\documentclass{aa}  

\usepackage{graphicx}
\usepackage{txfonts}
\usepackage{threeparttable}
\usepackage{hyperref}
\hypersetup{colorlinks,linkcolor={blue},citecolor={blue},urlcolor={blue}} 
\usepackage[normalem]{ulem}

\begin{document}

   \title{Particle acceleration and multi-messenger radiation from ultra-luminous X-ray sources:}
   \subtitle{A new class of Galactic PeVatrons}

   \author{Enrico Peretti
          \inst{1}\fnmsep\thanks{enrico.peretti.science@gmail.com}
          \and
          Maria Petropoulou\inst{2}\fnmsep\inst{3}
          \and 
          Georgios Vasilopoulos\inst{2}\fnmsep\inst{3}
          \and
          Stefano Gabici\inst{1}
          }

   \institute{Université Paris Cité, CNRS, Astroparticule et Cosmologie, 10 Rue Alice Domon et Léonie Duquet, 75013 Paris, France
         \and
             Department of Physics, National and Kapodistrian University of Athens, University Campus Zografos, GR 15784, Athens, Greece
         \and
         Institute of Accelerating Systems \& Applications, University Campus Zografos, Athens, Greece
             }

   \date{Received ...; accepted ...}

  \abstract 
  {Super-Eddington accretion onto stellar-mass compact objects powers fast outflows in ultra-luminous X-ray sources (ULXs). Such outflows, which can reach mildly relativistic velocities, are often observed forming bubble structures. Wind bubbles are expected to develop strong wind termination shocks, which are sites of great interest for diffusive shock acceleration. We developed a model of diffusive shock acceleration in the wind bubbles powered by ULXs. We find that the maximum energy in these objects can easily reach the PeV range, promoting winds from ULXs as a new class of PeVatrons. We specialized our model in the context of the Galactic source SS~433 and show that high-energy protons in the bubble might explain the highest energy photons ($>100$ TeV) and their morphology recently observed by LHAASO. In this paper, we discuss the detectability of such a source in neutrinos, and we analyze the possible radio counterpart of ULXs focusing on the case of W50, the nebula surrounding SS~433. 
  Finally, we discuss the possible contribution of Galactic ULXs to the cosmic-ray flux at the knee, concluding that their role could be significant only if one of these sources, currently undetected, were sufficiently close.}   

   \keywords{{\bf Acceleration of particles -- Astroparticle physics -- Cosmic rays}
               }

   \maketitle

\section{Introduction}

Ultra-luminous X-ray sources \citep[ULXs; see review][]{King2023} are loosely defined as point-like sources located away from the galactic nucleus and characterized by fluxes, which, assuming isotropic emission, translate to an X-ray luminosity $L_X\gtrsim 10^{39} \, \rm erg \, s^{-1}$. Over the past decade our understanding of ULXs has vastly expanded, with almost 2000 ULXs having been identified thus far \citep[][]{2022MNRAS.509.1587W}. Moreover, after the discovery of pulsating ULXs \citep[e.g.,][]{2014Natur.514..202B,Israel2017b,Amato2023} and various spectral studies \citep[e.g.,][]{Pintore2014,2017A&A...608A..47K}, it is now widely accepted that ULX engines are fueled by super-Eddington accretion onto stellar-mass compact objects~\citep{Israel2017}, instead of sub-Eddington accretion onto intermediate-mass black holes as early studies suggested \citep{1999ApJ...519...89C}. 

Extreme accretion goes hand in hand with strong winds and outflows~\citep{Fabrika2015,Kaaret2017,Weng2018}.
Outflows have been seen in Galactic X-ray binaries hosting both neutron stars and black holes accreting below the Eddington limit \citep[e.g.,][]{2012MNRAS.422L..11P,2023Natur.615...45V}. However, perhaps one of the early arguments for super-Eddington accretion came from the studies of a low-mass X-ray binary, Cyg~X-2. \citet{1999MNRAS.309..253K} argued that the system had survived a phase of extreme accretion ($10^{-6}-10^{-5}\, M_{\odot}\,\rm  yr^{-1}$) during which about only $0.1 M_{\odot}$ was accreted and $2-3~M_{\odot}$ were expelled, suggesting that Cyg~X-2 had once gone through a ULX phase.  
For ULXs, observing campaigns with the Reflection Grating Spectrometer of {XMM-Newton} revealed the presence of blueshifted absorption features consistent with outflow velocities of $0.2-0.3 \, c$ \citep[e.g.,][]{2017MNRAS.468.2865P,2020MNRAS.491.5702P}. 
It should also be noted that the presence of these winds and their properties are also supported by simulations \citep{Kitaki2021}.
These radiative-driven outflows are optically thick and can reprocess the accretion luminosity, while due to their inhomogeneity they imprint variability onto the X-ray emission and spectral hardness \citep[][]{2015MNRAS.447.3243M,2021A&A...654A..10G}. 
The basic idea is that these optically thick outflows block escaping radiation along certain lines of sight, allowing a distant observer to only view the central engine along a narrow funnel. This geometric structure would cause collimation of the radiation -- commonly referred to as beaming -- along the funnel and block radiation for an edge-on observer. Since beaming enhancement would be proportional to mass-accretion rate \citep{2009MNRAS.393L..41K}, the majority of ULXs could perhaps be hidden from us due to their unfavorable orientation.

One of those hidden ULXs is the Galactic source SS\,433, where studies have argued that the central engine accretes at a rate of $10^{-5}-10^{-4} \, M_{\odot} \, \rm yr^{-1}$ \citep{1981SvA....25..315S}. The properties of the outflows in SS\,433 were recently studied by \citet{2021MNRAS.506.1045M}. The authors modeled broadband X-ray spectra from {NuSTAR} and demonstrated that emission is consistent with reflection from the outflow funnel walls with an outflow velocity of $0.15-0.3 \, c$ and a wind-cone half-opening angle of 10 degrees.

Cygnus X-3 was recently characterized as another hidden ULX by IXPE observations \citep{Valedina_Cygnus_X3}. Specifically, the authors found that the energy-independent linear polarization is orthogonal to the direction of the radio ejections, indicating the presence of a collimating outflow with a half-opening angle less than 15 degrees. Yet, subsequent studies have also suggested that a face-on observer would observe Cygnus X-3 with an apparent luminosity of $5 \cdot 10^{39} \rm erg \, s^{-1}$, making the system a bona fide ULX \citep{Valedina24}.

Quasi-spherical, shocked ionized structures are typically observed in association with ULXs \citep[see, e.g.,][]{Pakull2002,Soria2021,Gurpide2022}. 
These structures can extend for more than $10^2$ pc in diameter, implying a million year lifetime of the associated engine. 
These observations suggest that ULX-driven winds not only result in relatively collimated jets, but typically they expand with a wide opening angle forming a bubble structure \citep{Belfiore2020,Zhou2022,Zhou2023}.
Wind bubbles are extremely interesting objects in the context of high-energy astrophysics as they can provide an ideal environment for diffusive shock acceleration (DSA). 
In fact, the evolution of wind bubbles results in the formation of strong shocks \citep{Weaver1977,Koo1992}, where cosmic rays (CRs) can be accelerated. 
In the context of DSA, the high velocity of ULX winds is of particular interest as the acceleration timescale of CRs is expected to scale as the inverse of the second power of the wind speed~\citep{Drury1983}. 
This would make ULX winds extremely efficient accelerators up to the PeV range, which is believed to be the range of maximum energy for Galactic CR sources~\citep{Blasi2013,Gabici2019}.
The nature of the Galactic CR accelerators at PeV energies, often referred to as PeVatrons~\citep{Cristofari2021}, is one of the main open questions in the CR community that has not been answered yet. 
However, recent gamma-ray observations performed by LHAASO~\citep{LHAASO_catalog} identified several Galactic sources whose non-thermal radiation clearly implies the presence of PeV particles. 
{Some of these sources coincide with Galactic super-Eddington accreting
objects~\citep{LHAASO_BH_2024}, which could be "hidden" ULXs. This supports the
hypothesis that super-Eddington sources are a new class of PeVatrons.}

Fast winds from stellar-mass compact objects and microquasars were known as sites of non-thermal emission~\citep[see, e.g.,][]{Romero2005,Escobar:2021ytm,Escobar:2022jod,Abaroa2023}.
{In addition, \cite{Abaroa2024}, studying the timescales of relevant high-energy processes, already identified microquasar jets powered by super-Eddington accretion as a promising site for proton acceleration up to the PeV range  \citetext{see also \citealp{Sridhar:2022uis} for similar implications in super-Eddington hypernebulae}.}

In this work, we developed a model of DSA taking place in the wind bubbles excavated by ULXs. We did so by solving the energy-dependent and space-dependent transport equation and analyzing the possible multi-messenger implications in terms of CRs, gamma rays, radio, and high-energy neutrinos.
Our investigation shows that ULX wind bubbles are extremely promising cosmic accelerators that in several conditions can energize CR protons up to the PeV range.
Therefore, we propose Galactic ULXs as a novel class of PeVatron candidates through DSA.

The manuscript is organized as follows. 
In Sect.~\ref{sec:model}, we outline the ULX wind-bubble structure and the acceleration and transport model for non-thermal particles. 
In Sect.~\ref{sec:results}, we present our results in terms of accelerated particle spectra and maximum energies achieved at ULX wind shocks. 
In Sect.~\ref{Subs: SS433}, we describe how we applied our model to SS~433, a hidden Galactic ULX, and discuss our predictions in light of recent gamma-ray observations at very high energies. We also analyzed the possible neutrino and radio counterparts of our ULX bubble model.
In Sect.~\ref{sec:pevatrons}, we discuss the contribution of Galactic ULXs to the observed cosmic-ray flux on Earth at energies close to the knee of the spectrum. Finally, we conclude in Sect.~\ref{sec:conclusions}.

\section{Model}\label{sec:model}

{When a compact object accretes mass from a nearby companion, the accretion flow is typically characterized by a certain angular momentum. 
This results in the formation of an accretion disk. 
Due to magnetic and viscous forces between disk annuli, there is an outwards transfer of angular momentum and the disk material approaches the compact object \citep{Shakura1973,Shakura_1976}. 
If the accretion rate exceeds the Eddington limit, $\Dot{M}_{\rm Edd} = L_{\rm Edd}/\eta_{\rm rad} c^2$, where $L_{\rm Edd}$ is the Eddington luminosity and $\eta_{\rm rad}(\sim0.1)$ is the energy conversion efficiency, the radiation pressure in the innermost part of the disk exceeds the thermal pressure and the disk inflates. 
In particular, the disk thickness grows linearly with the radius, thereby developing a funnel-like structure.
The accretion rate in the innermost region of the disk is bounded at the Eddington rate, while the excess mass flux is ejected in the form of a fast outflow \citetext{see e.g. \citealp{Abaroa2023}, \citealp{Abaroa_Mex2024}, and references therein}.
{The formation of winds from super-Eddington (or supercritical)
accretion was hypothesized by~\cite{Shakura1973} and thoroughly investigated in
pioneering work in the 1970s \citep{Meier1,Meier2,Meier3,Meier4}.}
Recently, progress in numerical simulations allowed a more detailed comprehension of the wind dynamics such as the typical launching radius, the angular properties of the outflow, and the connection of the kinetic power to the X-ray luminosity \citep{Kitaki2021}.
}

Ultra-luminous X-ray source activity may result in many sites of interest for particle acceleration and non-thermal processes, {such as shocks developed nearby the wind launching radius or other shocks developed at relatively small scales due to the collision between the wind of the donor star and the outflow from the accretion disk \citep{2022MNRAS.509.2532C, Abaroa2023}}. 
{In addition, the accretion onto the compact object might also result in the launching of a highly collimated jet reaching at least mildly relativistic velocities \citep{Blandford1,Blandford2}. As discussed in \cite{Globus16}, the existence of wide-opening-angle disk winds and jets, which are typically associated {with} micro-quasar activity, is not mutually exclusive. 
Besides, the jet itself is also a relevant site for particle acceleration and emission of non-thermal radiation \citep{Hess_SS433,Hawk_SS433,Safi-Harb2022}.}

In this work, we focused on {the} wind-blown bubbles {inflated by the disk winds,} and particularly on the DSA scenario at their wind termination shocks {that usually take place at tens of parsecs from the central engine}. 
{The simultaneous presence of a disk wind and a collimated jet, even if possible, as attested in the case of SS433 \citep{Watson1983,2021MNRAS.506.1045M}, can be expected to deform and elongate the structure of the bubble along the jet axis, as observed in the case of W50 \citep{Churazov}. 
In our work, we neglected the possible interactions between co-existing disk winds and jets, which may lead to additional sites of particle acceleration, and we focused on the wind bubble on its own in order to understand its multi-messenger signatures.}

\begin{figure}
    \centering
    \includegraphics[width=0.9\columnwidth]{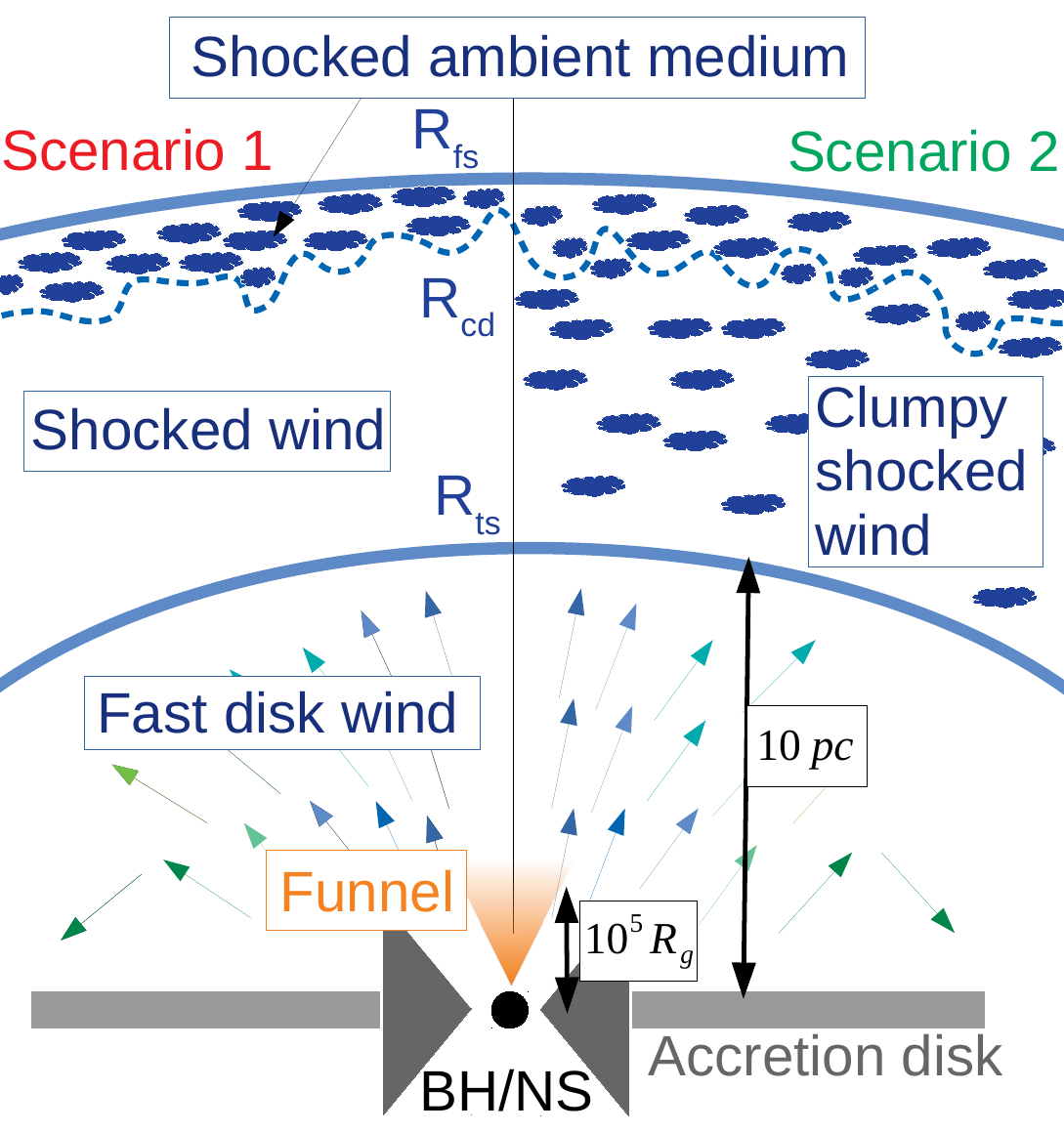}
    \caption{ULX sketch showing various components (not to scale) that are relevant to our model. Two scenarios for the density distribution in the shocked wind  that are investigated are displayed.}
    \label{Fig: ULX sketch}
\end{figure}

Figure~\ref{Fig: ULX sketch} illustrates a not-to-scale representation of the system where we highlight the key ingredients of a ULX and the associated wind bubble where particle acceleration and transport take place. 
A compact object such as a stellar-mass black hole (BH) or a neutron star (NS) in super-Eddington accretion develops a puffed part in the innermost region of the accretion disk\footnote{{While the structure of a super-Eddington accreting disk may differ from the one shown in the sketch \citep[according to][the accretion disk might keep a constant height at large distances from the black hole, yet at small radii one would have the funnel-like behavior typical of super-Eddington systems.]{Kitaki2021}, this does not affect the large-scale structure and evolution of the outflows.}}, while a baryon-loaded fast wind with wide opening angle is launched at high speed.
Such a wind, as described in Sect.~\ref{Subs: wind-bubble}, is a complex, stratified structure characterized by a strong shock surrounding a fast-wind region. 
The latter is then surrounded by a hot and extended shell of shocked wind material and by an additional outer thin-shell of swept-up shocked ambient medium.
Particles can be accelerated by DSA at such a strong shock and being advected and diffusing throughout the whole bubble. 
In Sect.~\ref{Subs: Transport}, we highlight the details of the model. 

\subsection{Wind bubble structure}
\label{Subs: wind-bubble}

We worked under the assumption that the ULX wind blows approximately steadily in time, with a wide opening angle in a spatially uniform interstellar medium (ISM), thereby developing a bubble structure. A wind bubble forms as follows.
At a given time $t_0,$ the ULX starts blowing a fast wind with wide opening angle. 
The wind is supersonic; consequently, it drives a forward shock (FS) expanding into the external medium. 
During the expansion the wind sweeps up the external medium, which gets shocked and accumulated in a thin layer behind the FS. Such a layer is separated from the wind material by the contact discontinuity (CD), the physical boundary of the wind. 
As the system expands, the collision of the wind with the external medium results in the formation of a second shock inside the wind structure, known as wind termination shock (TS), which has its upstream region facing the central engine.
We note that, in the reference frame of the central engine, the TS expands radially as the FS does, whereas it travels backward (as a reverse shock) only as seen from the reference frame of the fast wind.   
After an initial phase of free expansion with velocity $V_w$, as soon as the swept-up external matter becomes roughly comparable with the wind mass, the system enters the deceleration phase. 
This happens roughly on a timescale of $t_{\rm dec} \approx [3 \,  \Dot{M}/(4 \pi m_p n_{\rm ext} V_w^3 )]^{1/2} \approx 3~\rm yr \,  \Dot{M}^{1/2}_{-6} \, n_{\rm ext, 0}^{-1/2} \, V_{w,9}^{-3/2}$, where $\Dot{M}$ ($\Dot{M}_{-6}$) is the wind mass-loss rate (in units of $10^{-6} M_{\odot} \rm yr^{-1}$), $n_{\rm ext}$ ($n_{\rm ext, 0}$) is the density of the external medium (in units of $\rm cm^{-3}$), $V_{w,9}$ is the terminal wind speed in units of $10^9 \, \rm cm \, s^{-1}$, and $m_p$ is the proton mass.
In this phase, the FS and the TS expand self-similarly with scaling $R_{\rm fs}\sim t^{3/5}$ and $R_{\rm ts}\sim t^{2/5}$ \citep{Weaver1977}.
In particular, the temporal evolution of the shocks can be written as follows~\citep{Koo1992}:
\begin{align}
\label{Eq. FS}
        R_{\rm fs}(t) & \approx 120 \, {\rm pc} \, t_6^{3/5} (\Dot{E}_{39} \, n_{\rm ext, 0}^{-1})^{2/10} ,\\ 
        R_{\rm ts}(t) & \approx  14.5 \, {\rm pc} \, t_6^{2/5} \, (\Dot{E}_{39}\,n_{\rm ext, 0}^{-1})^{3/10} V_{w,9,}^{-1/2} 
 \label{Eq. TS}
\end{align}
where $t_6$ represents the time in units of Myr and $\Dot{E}_{39}$ is the kinetic power of the outflow ($\Dot{M}V_w^2/2$) expressed in units of $10^{39} \, \rm erg \, s^{-1}$.

As one can see from Eq.~\eqref{Eq. FS}, the FS rapidly slows down in the ISM. 
Such a shock is also expected to become radiative as the shocked ambient medium (SAM) typically has a short cooling time.
Eq.~\eqref{Eq. TS} implies that the TS slows down even faster than the FS, possibly stalling in the reference frame of the central engine.
However, as the innermost fast wind launched by the ULX and cooled adiabatically impacts on it at a very high velocity, such a shock is expected to be a strong shock.

The late stage of the bubble, $t \gg t_{\rm dec}$, is extremely interesting for particle acceleration and transport as the two shocks are evolving very slowly in time, while strong shock conditions are present at the TS. 
Therefore, DSA can take place efficiently at such a shock, while, approximately, the bubble can be assumed as a steady system.
On the other hand, the FS, being radiative and slowing down, is not expected to be an efficient accelerator~\citep[see, e.g.,][and references therein]{del_Valle22}. 
{The proper motion of the source might result in the distortion of the bubble shape in the late deceleration~\citep{Bosch-Ramon2011-pm}. While this can impact the source morphology, it is unlikely to produce an observable impact on the gamma-ray spectrum. Therefore, we leave it for a follow-up investigation.}

To sum up, the bubble is a spherically symmetric onion-like structure characterized by the three typical discontinuities: the innermost TS ($R_{\rm ts}$) separating fast cool wind from the hot shocked wind, the CD ($R_{\rm cd}$) separating the hot shocked wind from the SAM, and the outermost FS ($R_{\rm fs}$). 

The wind bubble is assumed to be energy-conserving, namely in conditions such that the shocked wind does not radiate. 
Thus, the wind velocity is $u(r) \propto r^{-2}$ for $r>R_{\rm ts}$. 
In the innermost region, for $r<R_{\rm ts}$ the wind velocity is with good approximation constant, with $u(r)=V_w$.
The number density profile is then computed from the mass continuity equation, given the wind velocity profile, as $n_1(r) = \Dot{M}/[4 \pi m_p r^2 V_{w}]$ for $r<R_{\rm ts}$, where for simplicity we have assumed a composition dominated by ionized hydrogen.
For $r>R_{\rm ts}$ we explored two alternative scenarios: 
\begin{itemize}
    \item S1: The standard thin-shell approximation in which the shocked wind is perfectly separated from the SAM, as represented by the small clumps in Fig.~\ref{Fig: ULX sketch} (left hand side), by the CD. In this context, the density of the shocked wind would be constant and equal to $n_2 = 4 \, n_{1}(R_{\rm ts})$.
    \item S2: The target material is uniformly distributed throughout the shocked wind region (thick shell), as represented in the same figure (right hand side) by the uniformly distributed clumps in the shell extending from $R_{\rm ts}$ up to $R_{\rm fs}$. This scenario of clumpy shocked wind (CSW), in which $n_2$ is a free parameter, can be the result of several dense molecular ISM clumps swallowed by the system, partial evaporation of the SAM~\citep{Weaver1977}, or the consequence of instabilities taking place at the contact discontinuity and allowing a sizable fraction of the SAM material to be flooded over by the shocked wind~\citep{Bosch-Ramon2010}. 
\end{itemize}

\subsection{Particle acceleration and transport in the bubble}
\label{Subs: Transport}

We focused on the late-time stages of the wind-bubble evolution, where the shock velocities in the reference frame of the central engine are low compared to $V_w$. 
In this context, the particle transport can be approximated as stationary, and the only relevant site for DSA is the TS. 
On the other hand, at the FS, particles are assumed to freely escape without experiencing any re-acceleration. 
As mentioned previously, this assumption is justified since the FS is expected to be rapidly slowing down and radiative~\citep{Drury1983}.

We assume spherical symmetry. {We point out that such an assumption is unjustified in the vicinity of the compact object due to angular dependence of the outflow velocity \citep{Kitaki2021} and the vicinity of the donor companion star \citep{Abaroa2023}. However, on parsec scales from the launching region it is reasonable to expect these small-scale irregularities to play a negligible role in the overall geometry of the bubble~\citep{Menegazzi:2024tzl}. 
{Nonetheless,} these spatial scales might be potentially interesting for the injection of mechanical turbulence in the fast cold wind. However, a deep investigation of the turbulence properties goes beyond the scope of this work.} 
We modeled particle acceleration and transport of high-energy protons in the wind bubble with the following stationary transport equation~\citep{Morlino2021,Peretti2022,Peretti2023,Payel2023}:
\begin{equation}
\label{Eq: transport}
     \nabla \cdot [ D \nabla f] - u \cdot \nabla f  + \frac{1}{3} p \frac{\partial f}{\partial p} \nabla \cdot u - \frac{f}{\tau_{\rm loss}} + Q = \frac{\partial f}{\partial t} = 0,
\end{equation}
where $f=f(r,p)={\rm d}N/{\rm d}^3p{\rm d}V$ is the particle distribution function depending on radius $r$ and momentum $p$. $D(r,p)$ is the diffusion coefficient, $u(r)$ is the velocity profile, and $\tau_{\rm loss}(r,p)$ is the energy-loss timescale for $pp$ interactions. {The latter process can be approximated by a catastrophic energy-loss term (i.e., particles of momentum $p$ are removed from the system on a timescale $\tau_{\rm loss}$ instead of moving to lower energies), since the proton in one collision typically loses a significant
fraction ($\kappa_{pp}\approx 0.5$) of its energy \citep[e.g.,][]{Kelner_pp}.} 
We stress that the stationary approximation is well justified, since in the late deceleration non-thermal particles evolve on timescales shorter than the dynamical time of the system.  
The injection term $Q$ has the following expression:
\begin{equation}
    \label{Eq: injection}
    Q = \eta \frac{n_1(R_{\rm ts})V_w}{4 \pi p^2} \delta[p-p_{\rm inj}] \delta[r-R_{\rm ts}],
\end{equation}
where $\eta$ is the fraction of the flux of all plasma protons arriving at the TS, $n_1(R_{\rm ts})V_w$, that is injected in the DSA. The two delta functions state that only particles at the wind termination shock position and with a certain momentum $p_{\rm inj}$ are accelerated. 
We note that $p_{\rm inj}$ is generally located in the high-energy tail of the thermal Maxwellian of the plasma, and it does not play any role on the CR pressure provided that it is not assumed to be much larger than $m_p c$. We assume $p_{\rm inj} = m_p c$.
{Finally, we notice that {\cite{Takeuchi2013} and \cite{Kobayashi2018}} pointed out that ULX winds might be characterized by a substantial clumpiness, especially close to the launching region. 
As they are not expected to have a large filling factor over the whole multi-parsec-sized volume, their impact on the transport of particles in the wind bubble is not expected to have an observable impact.}

Our model focuses on the hadronic acceleration and radiation, while we limit our discussion of primary electrons and their radio emission to Sect.~\ref{Subs: SS433}.
We also leave the inclusion of the secondary electron-positron populations and their multiwavelength emission to future investigations, as the secondary-to-primary electrons ratio is expected to be smaller than unity for ULXs in the energy range of interest for radio emission.
It is worth mentioning that, as the inferred magnetic field in these bubbles is on the order of a few $10 \, \rm \mu G$, the synchrotron cooling of electrons is at least an order of magnitude more efficient than the inverse Compton cooling on the cosmic microwave background. 
Consequently, one can expect the leptonic emission to be relevant in radio and X-rays while being subdominant in the gamma-ray domain. 

We estimated the magnetic-field pressure ($B^2/8\pi$) in the fast cool wind as a fraction, $\epsilon_B$, of the wind kinetic energy density ($\rho V_w^2/2$).
We computed the magnetic-field amplitude in the downstream region, assuming that only the components perpendicular to the shock normal are compressed, such that the downstream magnetic field is amplified by a factor $\sqrt{11}$ with respect to the upstream one~\citep{Marcowith2010}. 
The downstream magnetic-field amplitude reads as follows:
\begin{equation}
B = 8.5 \, \mu {\rm G} \,  \epsilon_{B,-1}^{1/2} \Dot{M}_{-6}^{1/2} V_{w,9}^{1/2} R_{\rm ts,1}^{-1},
\label{eq:B}
\end{equation}
where $R_{\rm ts,1}$ is {the TS radius} normalized to 10~pc.
Such a magnetic field is assumed spatially constant in the downstream.
CRs are scattered off magnetic irregularities, such as Alfv{\' e}n waves. The associated diffusion coefficient is computed in the context of the quasi-linear theory as \citep{Blasi2013}
\begin{equation}
\label{Eq: Diff}
    D(r,p) \approx \frac{v(p)}{3} 
    \begin{cases}
    r_L^{2-\delta}{(r,p)} \, l_c^{\delta-1}, & r_L \leq l_c  \\
    r_L^2{(r,p)} \, l_c^{-1}, & r_L \geq l_c
    \end{cases} 
,\end{equation}
where $v(p)$ is the particle velocity, $r_L$ is the Larmor radius, $l_c$ is the coherence length of the magnetic field, and $\delta$ is the spectral index of the turbulence cascade. 
We highlight that the diffusion coefficient is assumed to scale as $r_L^2$ when $r_L \gtrsim l_c$~\citep{Subedi2017}.
In this work, we mostly focused on $\delta=3/2$ as prescribed for magneto-hydrodynamical turbulence \citep{Kraichnan}. However, in Sect.~\ref{Subsub: analytic emax}, we develop our analytic discussion in the context of Bohm diffusion, $\delta=1,$ in order to provide some general upper limits to the maximum energy of accelerated particles.

We solve Eq.~\eqref{Eq: transport} following the semi-analytic approach developed in \cite{Morlino2021} and \cite{Peretti2022,Peretti2023}. This is based on solving Eq.~\eqref{Eq: transport} in the upstream and downstream regions separately and finally joining the two spatial solutions at the wind-shock location. We refer the interested reader to the above-mentioned works for additional information on the solving algorithm.  

As it is useful for qualitative understanding, we report the form of the solution at the wind TS:
\begin{equation}
    f_{\rm ts}(p) = C p^{-s} e^{-\Gamma_1(p)} e^{-\Gamma_2(p)},
\end{equation}
where $s \approx 4 $ is the power-law index, while the $\Gamma_i$ (i=1, 2) are two monotonic functions of momentum that set the maximum energy due to the finite size of the upstream and downstream regions. The normalization factor $C$ is computed assuming that the pressure of the accelerated particles is a fraction  $\xi_{\rm CR} \lesssim 0.1$ of the wind ram pressure at the TS so that the shock structure would not be modified by the CR pressure (test particle approximation). 
In general, for fixed $\xi_{\rm CR}$ faster winds {(with the same $\Dot{M}$)} favor higher cosmic-ray pressures at the TS.
Clearly, $\xi_{\rm CR}$ and the parameter $\eta$ in Eq.~\eqref{Eq: injection} are different parametrizations of the same physical quantity. Hereafter, we limit our discussion to $\xi_{\rm CR}$ as its meaning is more intuitive.

{We computed the production spectra of gamma rays
following the approach described in \cite{Kelner_pp}. 
We report the details of this calculation in Appendix~\ref{Appendix}.}

Finally, as it can be useful for general considerations of the contribution to the observed CRs, we report the expression of the escaping flux of particles from the FS of a ULX in the following \citep[see also][]{Peretti2022,Peretti2023}:
\begin{equation}
\label{Eq. esc flux}
    J_{\rm esc}(p) = \frac{{\rm d}N}{{\rm d}^3p \, {\rm d}t} = \frac{\pi \, R_{\rm ts}^2 V_{w} \, [1 - \mathcal{L}(p)]}{1 - \exp\{-R_{\rm ts}V_{w}(1-\frac{R_{\rm ts}}{R_{\rm fs}})/[4 D_2(p)]\} } f_{\rm ts}(p),
\end{equation}
where the function $\mathcal{L}$ accounts for the reduction to the escaping flux due to the $pp$  catastrophic energy losses.  
As $pp$ losses have a negligible observational impact on the bulk of accelerated particles, one generally expects $\mathcal{L} \ll 1$.

\section{Results}\label{sec:results}
\renewcommand{\arraystretch}{1.1}
\begin{table}
\centering
\begin{threeparttable}
\caption{Parameters for wind bubble of a prototypical ULX.}
\begin{tabular}{c c c}
    \hline
     Parameter & Value & {Description} \\
     \hline
     $V_{w}$ & $0.1c -0.2 \, c$ & {terminal wind speed} \\
     $\Dot{M}$ & $10^{-6} \, \rm M_{\odot} \, yr^{-1}$ & {mass loss rate} \\
     $\frac{\epsilon_B}{2}$ & $0.1$ & {mag. to ram pressure ratio} \\
     $l_c$ & $0.1 \, \rm pc$ & {coherence length} \\
     \hline
     $\tau_{\rm age}$ & $2 \cdot 10^{5} \, \rm yr$ & {age of the ULX} \\
     $n_{\rm ext}$ & $1 \, \rm cm^{-3}$ & {external density} \\
     
     \hline
    \label{Table: SS433}
\end{tabular}
\end{threeparttable}
\end{table}

In what follows, we discuss the results of our model.   
We remind the reader that, as we are mostly interested in the gamma-ray domain, we regard wind bubbles as {hadronic-only} sources. 
In Sect.~\ref{Subs: particles}, we discuss the spectra of non-thermal accelerated particles characterizing ULXs, while in Sect.~\ref{Subsec: E_MAX}, we focus on the maximum energy achievable by non-thermal particles in these objects. 

\subsection{Timescales and spectra of high-energy particles}
\label{Subs: particles}
 
We studied the non-thermal hadronic signatures of ULX bubbles by selecting a set of representative parameters for a prototype source (see Table~\ref{Table: SS433}). 

\begin{figure}[t!]
    \centering
    \includegraphics[width=1.\columnwidth]{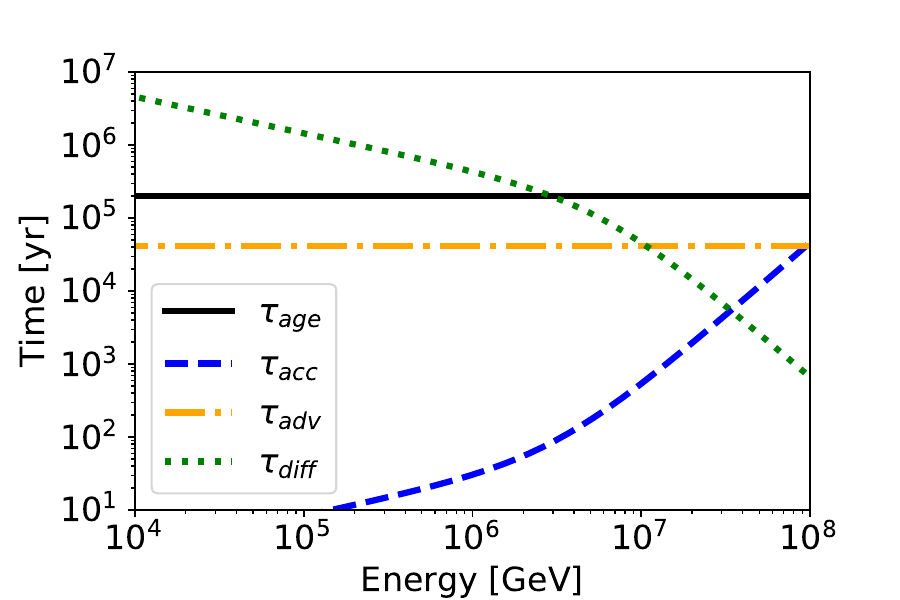}
    \caption{Typical timescales regulating behavior of non-thermal particles in ULX wind bubble. The acceleration timescale (blue dashed) is compared with the timescales of the main escape processes: advection (yellow dot-dashed) and diffusion (green dotted). The age of the system is also shown as a solid black line.}
    \label{Fig: Timescales bubble}
\end{figure}

Even though the typical timescales of the main physical mechanisms do not have a direct mathematical impact on the solution of our stationary model, they are extremely useful in guiding a qualitative understanding of the dominant processes in different energy ranges. Figure~\ref{Fig: Timescales bubble} illustrates the comparison of the timescales of the main physical processes taking place at the wind TS. The age of the system (solid black line) is much longer than the acceleration timescale ($\tau_{\rm acc}\approx s D_1/V_{w}^2$), which is represented in the plot by the dashed blue line, and always subdominant with respect to the shortest escape time. 
At low energies, the advection of particles from the TS -- which happens on a timescale of $\tau_{\rm adv}=(R_{\rm fs}-R_{\rm ts})/\langle u_2 \rangle$ -- is the dominant escape mechanism, whereas at energies of $\gtrsim 3 \cdot 10^7~\rm GeV$, the diffusion, whose characteristic timescale is $\tau_{\rm diff}=(R_{\rm fs}-R_{\rm ts})^2/D_2$, takes over and sets an upper limit to the maximum energy. Here, the subscript "2" states that the quantity is computed in the downstream region of the TS. 
{We do not show the adiabatic timescale as, for a downstream wind profile of $\sim r^{-2}$, adiabatic energy losses do not play a role in the downstream region.}
We finally note that the $pp$ loss timescale, $\tau_{pp} \approx (n \kappa_{pp} \sigma_{pp}c)^{-1}$, is not included in the figure, as it is many orders of magnitude longer than all other timescales. For example, in the context of the thin-shell approximation, the density of the shocked wind can be written as $n_{\rm 2} \approx 10^{-5} \, \Dot{M}_{-6}^{-1} \, V_{w,9} \, R_{\rm ts,1}^2 \, \rm cm^{-3}$, resulting in a loss timescale longer than the age of the Universe. Assuming a higher density, as prescribed for S2, would lead to timescales of the order of 5-50 Myr. Only the density of the SAM layer in S1 might provide a reasonably short timescale. However, the residency time of particles in such a small layer would be reduced by some orders of magnitude due to its reduced radial extension.
This means that the ULX bubble is far from being a hadronic calorimeter.

\begin{figure}[t!]
    \centering
    \includegraphics[width=1.\columnwidth]{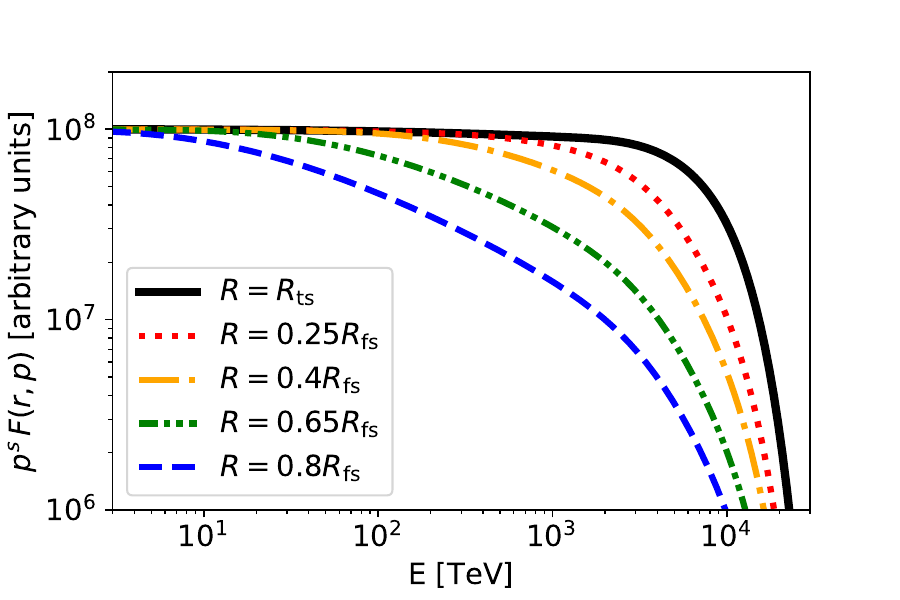}
    \caption{Proton spectra at different radii. The spectrum at the TS (solid black) is compared with the spectra at progressively larger radii approaching the FS (from the red dotted to the blue dashed).}
    \label{Fig: Radial behavior}
\end{figure}

Figure~\ref{Fig: Radial behavior} illustrates the spectral shape of the proton distribution as computed at different radii throughout the downstream region. 
In particular, the solid black line represents the solution at the shock, while the other line styles and colors represent the proton spectra at progressively larger radii starting from the TS and moving outward to the FS. 
One can immediately notice that, as suggested from the timescales, the maximum energy\footnote{The maximum energy, following \cite{Morlino2021}, is defined as the energy where $p^sf$ drops by one e-folding compared to its low-energy asymptotic value.} at the shock surpasses the PeV range, reaching the value of $E_{\rm max} \approx 8 \, \rm PeV$.
The radial dependence of the spectrum shows the impact of diffusion, which becomes more and more relevant at larger radii. 
Finally, as the energy losses in the system are subdominant, the escape spectrum is not substantially different than the spectrum of accelerated particles at the TS.

\subsection{Maximum energy in ULX wind bubbles}
\label{Subsec: E_MAX}

The maximum energy that particles can reach at a cosmic accelerator is a fundamental property in the context of the origin of the highest energy both Galactic and extra-Galactic CRs. 
In what follows, we first provide an analytic argument aimed at understanding the maximum energy accessible to particles accelerated at wind termination shocks of different wind-blown bubbles depending on their main parameters. 
The qualitative discussion is then followed by a quantitative analysis of the model outcomes and a parameter-space scan.

\subsubsection{Analytic argument} 
\label{Subsub: analytic emax}

In order to support our claim of ULX wind bubbles being new Galactic PeVatron candidates, we provide a simple analytic argument resulting in a general upper limit. 
Differently from the rest of this manuscript, in this subsection we assume the simplest and most optimistic micro-physical conditions provided by the Bohm diffusion, $\delta = 1$. 

We derive a physically motivated upper limit on the maximum energy at the TS equating the upstream diffusion length $D_1/V_{w}$ to a sizable fraction $\chi$ of the TS radius $R_{\rm ts}$. This results in the following condition: 
\begin{equation}
\label{Eq: E_max}
    E_{\rm max} = \frac{3q}{c} \chi \epsilon_B^{1/2} \Dot{M}^{1/2} V_{w}^{3/2} \simeq 2.4~{\rm PeV} \, \chi \epsilon_{B,-1}^{1/2} \Dot{M}_{-6}^{1/2} V_{w,9}^{3/2}.
\end{equation}

\begin{figure}
    \centering
    \includegraphics[width=\columnwidth]{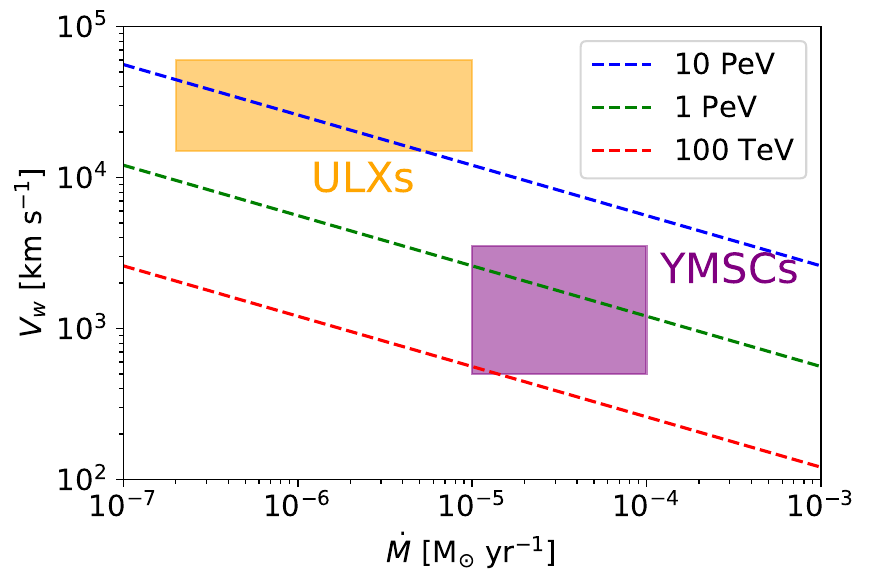}
    \caption{$\Dot{M}-V_{w}$ parameter space of typical Galactic wind bubbles and achievable maximum energy computed according to Eq.~\eqref{Eq: E_max}.}
    \label{Fig: Hillas Bubble}
\end{figure}

We highlight that while the maximum energy has a strong dependence on the wind speed, it is only mildly dependent on the mass-loss rate. 
Eq.~\eqref{Eq: E_max} depends on a few parameters of the system allowing a direct qualitative comparison with other Galactic sources characterized by wind bubbles, such as those inflated by young massive stellar clusters (YMSCs).
Figure~\ref{Fig: Hillas Bubble} illustrates the parametric dependence of the maximum energy, where we assume $\chi = 1$ and $\epsilon_B = 0.1$. 
Interestingly, differently from the standard Hillas plot where the maximum energy is set by the source size and magnetic field, here the parameter space can be reduced to $\Dot{M}$ and $V_{w}$, which are the main macro-physical parameters of these objects. 
The range of parameters shown in the figure are taken from \cite{King2023} and \cite{Stevens_YMSC2003} for ULXs and YMSCs, respectively.
The dashed lines identify three different maximum energies given by Eq.~\eqref{Eq: E_max}.

We remind the reader that these contours were computed assuming the most optimistic conditions for the bubble, set by the Bohm diffusion. As a result, Eq.~(\ref{Eq: E_max}) should be interpreted as an upper limit to the true maximum energy. Our results suggest that only the  most extreme YMSCs could possibly access the PeV range; whereas, because of their faster outflows, ULXs stand as an extremely promising class of Galactic PeVatrons.

\subsubsection{Parameter-space scan}
\label{Subsub: Emax-param-scan}

In our wind-bubble model, the power of the system is set by $\Dot{M}$ and $V_w$, while $\epsilon_B$ and $l_c$ are related to the properties of the magnetic field and affect the diffusion of particles.
While the mass-loss rate and terminal wind speed can be inferred from observations, the magnetic field is typically unknown, thereby making its associated parameters highly unconstrained.
In particular, the coherence length could extend from a small up to a sizable fraction of the size of the system.
Similarly, the magnetic-to-ram pressure ratio could range from $\sim10\%$ down to a few percent (and possibly lower values). For each of the four mentioned parameters, we set up a variability range and explored the resulting maximum energy while keeping the age of the system to 1~Myr constant.

Table~\ref{table: Emax-proxi} illustrates the impact of the four parameters on $E_{\rm max}$ (expressed in PeV). In particular, we explored three possible values of $l_c$ (0.01~pc, 0.1~pc, 1~pc) and $\epsilon_B$ (0.01, 0.03, 0.1) and two values of $\Dot{M}$ ($10^{-6}$-$10^{-5}$ $M_{\odot} \, \rm yr^{-1}$) and $V_w$ ($10^{4}$-$10^{5}$ $\rm km \, s^{-1}$). It is possible to observe that the qualitative trend expressed by Eq. \eqref{Eq: E_max} is approximately confirmed for the three parameters characterizing such relation, even though the diffusion scenario adopted here, $\delta=3/2$, is different from the one of the simple analytic argument leading to Eq.~\eqref{Eq: E_max}. 
In particular, in order to give a quantitative idea, the normalization at 10 GeV of the upstream diffusion coefficient at the shock can be written as $D_1(10 \, {\rm GeV}) \approx 2 \cdot 10^{25} \, {\rm cm^2 \, s^{-1}} \, R_{\rm ts,1}^{1/2} \, l_{\rm c,-1}^{1/2} \, \epsilon_{\rm B,-1}^{-1/4} \, \Dot{M}_{-6}^{-1/4} \, V_{w,9}^{-1/4}$, where the lengths are expressed in decades of parsecs.

Conversely to the other parameters, it is possible to notice a nontrivial dependence of the $E_{\rm max}$ on $l_c$, which seems to have a local maximum around $l_c\approx 10^{-1} \, \rm pc$ at low velocities. 
This behavior can be ascribed to a transition in the diffusion regime. 
In fact, in the majority of realizations, the maximum energy is already reached in the small pitch angle scattering regime, where $E_{\rm max} \sim l_c^{1/2}$. While, for $l_c \approx 1 \, \rm pc $ and $V_{w} = 10^4 \, \rm km \,  s^{-1}$, the maximum energy occurs in the resonant scattering branch of the diffusion coefficient where $E_{\rm max} \sim l_c^{({1-\delta})/({2-\delta})}$.

The parameter space scan combined with the previous analytic argument suggests that typical ULXs can reach the PeV range for a wide range of mass-loss rates and terminal wind speeds. 
In particular, the maximum energy is an increasing function of these macroscopic parameters. 
The magnetic-field amplitude, parametrized by $\epsilon_B$, also influences the maximum energy monotonically, whereas the dependence of the latter on the coherence length is nontrivial as the maximum energy might occur in different regimes of the diffusion coefficient. 
Finally, we noticed that particularly powerful ULXs could access several tens of PeV.

\section{Multi-messenger emission from ULXs: Application to SS~433}
\label{Subs: SS433}

The Galactic source SS~433 has been identified as a plausible misaligned (hidden) Galactic ULX~\citetext{see, e.g., \citealp{2006MNRAS.370..399B} and \citealp{King2023}} that is powered by a super-Eddington accretion onto a compact object~\citetext{most likely a black hole, as discussed in, e.g., \citealp{Poutanen2006} and \citealp{2021MNRAS.506.1045M}}. SS~433 has been observed in very high-energy gamma-rays ($E_{\gamma} \gtrsim 1~$TeV) by H.E.S.S.~\citep{Hess_SS433}, HAWC \citep{Hawk_SS433}, and more recently LHAASO \citep{LHAASO_BH_2024}. In this section, we present our model predictions for the case of SS~433 using the representative set of parameters listed in Table~\ref{Table: SS433} as inferred to be appropriate for such an object \citep{King2023}. 

\begin{table}
\caption{Maximum {proton energy (in PeV) attained at wind TS} assuming an age of 1 Myr.}
\label{table: Emax-proxi}
\centering             
\begin{tabular}{c|c|c|c|c}
\hline 
$l_{c,-2/-1/0}$ & $V_{w,4} \, \Dot{M}_{-6}$ &  $V_{w,4} \, \Dot{M}_{-5}$ & $V_{w,5} \, \Dot{M}_{-6}$ & $V_{w,5} \, \Dot{M}_{-5}$ \\
\hline 
  $\epsilon_B = 0.1$ & .4/.5/.07  & 1/2/.5 & 6/18/27  & 14/41/86 \\
  $\epsilon_B = 0.03$ & .2/.3/.04 & .6/1/.2 & 3/10/15 & 8/23/50  \\
  $\epsilon_B = 0.01$ & .1/.2/.02 & .3/.6/.1 & 1.5/4/6 & 4/13/28 \\ 
 \hline
\end{tabular}
\tablefoot{The table is organized as follows. Each cell contains three values corresponding to different assumptions in the coherence length, as indicated in the top left corner. Different rows report different values of $\epsilon_B$, while different columns report different combinations of wind speed and mass-loss rate.}
\end{table}

\begin{figure}
    \centering
    \includegraphics[width=1\columnwidth]{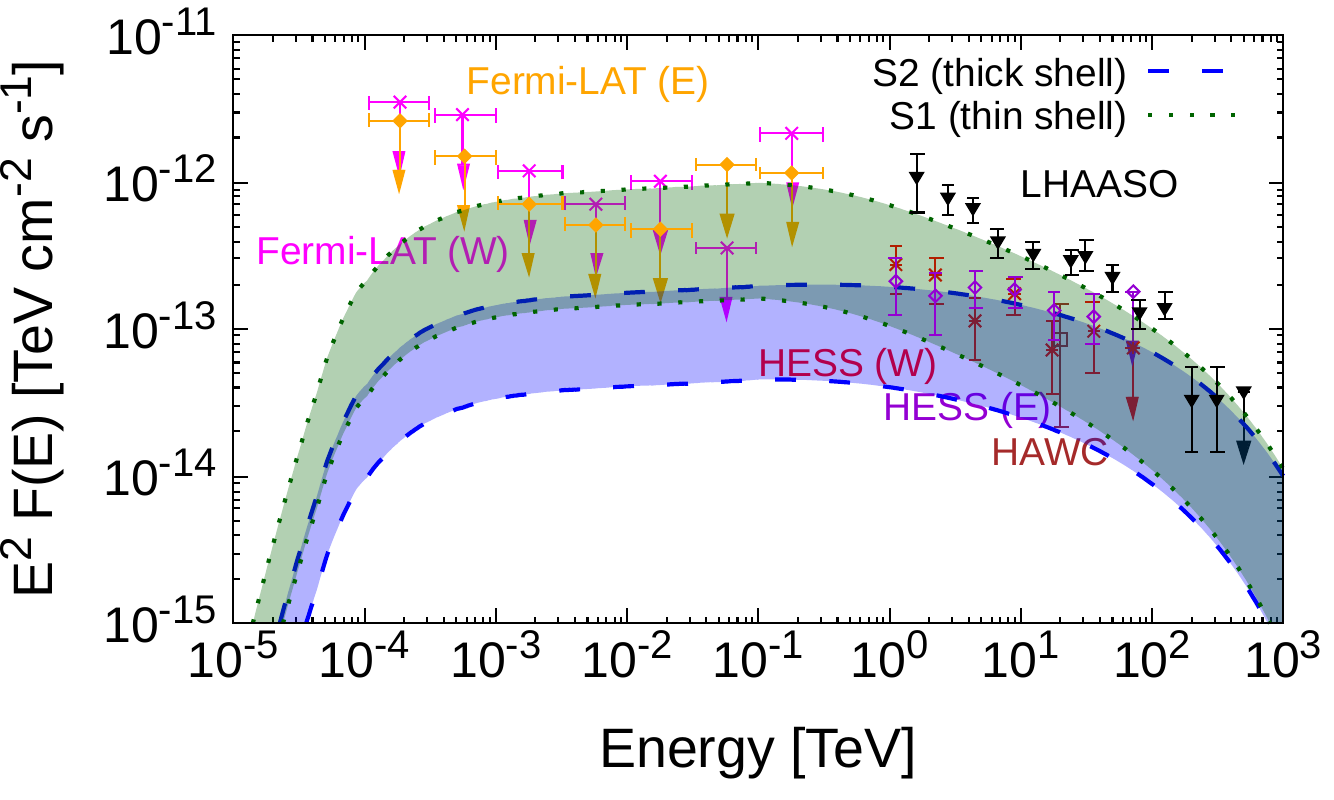}
    \caption{Spectral energy distribution of gamma-ray emission from SS~433.   Colored markers indicate flux measurements obtained with H.E.S.S. and HAWC. More recent data obtained by LHAASO are shown with black markers. The upper limits imposed by Fermi-LAT on the GeV emission from the eastern (E) and western (W) jets are overplotted with colored downward-pointing arrows. The predicted (hadronic) gamma-ray spectrum from the bubble is overplotted for the two scenarios presented in Fig.~\ref{Fig: ULX sketch}. The shaded region shows the expected range of gamma-ray emission when varying the terminal wind speed by a factor of two.
    }
    \label{Fig: Modelling SS433}
\end{figure}

As the energy losses are not expected to affect the proton distribution (see discussion about Fig.~\ref{Fig: Timescales bubble}), differences in the gamma-ray spectrum can only arise from different distributions of target material in the bubble for a fixed set of parameters. 
Fig.~\ref{Fig: Modelling SS433} illustrates the difference in the gamma-ray spectra obtained in the two scenarios, S1 (green curves) and S2 (blue curves), describing the downstream region of the TS (see sketch in Fig.~\ref{Fig: ULX sketch}). 
In particular, for S2 we assumed a downstream density of $n_{\rm CSW}= 0.1 \, \rm cm^{-3}$.
For each scenario the shaded area represents the model predictions obtained by varying the terminal wind speed from $0.1 \, c$ to $0.2 \,c$ while keeping the cosmic-ray efficiency at the shock, $\xi_{\rm CR}=0.1, $
constant. The corresponding maximum energy is found to range from 3 PeV up to 8 PeV.

The model predictions for the gamma-ray emission from the whole bubble were visually compared with HAWC~\citep{Hawk_SS433} and LHAASO~\citep{LHAASO_BH_2024} data, as well as the Fermi-LAT upper limits and H.E.S.S. data~\citep{Hess_SS433} collected from the eastern (E) and western (W) jets launched from the compact object powering the whole system. 
We note that as the model prediction is produced from the whole bubble embedding both jets, its contamination to the E and W lobes observed by H.E.S.S. should be scaled down by a factor accounting for the different angular extensions of the two regions. Consequently, the best impression of the actual contribution of the bubble to the overall emission can be deduced by comparing the model prediction with the LHAASO data. 

Interestingly, unless the most optimistic scenario is assumed, the wind-bubble model prediction can provide a dominant contribution -- of hadronic origin -- to the observed emission only above 100 TeV. 
It is intriguing that, in such a range, leptonic models are expected to fall more rapidly than the observed spectrum \citetext{see, e.g., Fig.~1 in \citealp{LHAASO_BH_2024}}. 
Moreover, the morphology of the gamma-ray source changes as a function of photon energy, with lower energy emission ($1-100$~TeV) following the jet geometry; whereas, the highest energy emission ($> 100$ TeV) appears to be more spherical \citep[see Fig.~1 in][]{LHAASO_BH_2024}. 
The angular extent of the $>100$~TeV emission translates to a minimum radius of $r_{39}\approx 30$~pc \citetext{taking the distance to the source to be 5.5 kpc in agreement with \citealp{2004ApJ...616L.159B} and \citealp{2007MNRAS.381..881L}}. Interestingly, using Eqs.~\eqref{Eq. FS} and \eqref{Eq. TS}, we find that $R_{\rm ts}\simeq 3$~pc and $R_{\rm fs}\simeq 35-47 \, \rm pc$ for the set of parameters assumed. 
{These results support the bubble interpretation and warrant further investigation.}

Finally, it is intriguing to notice that Fermi-LAT might soon be able to constrain the bubble emission in the GeV range from the hadronic scenarios discussed in this section. In particular, the detection of a relatively soft emission in the GeV band (softer than the rising part of an inverse Compton component) combined with a morphology similar to the LHAASO emission $>100$ TeV would be an additional strong indication supporting the bubble interpretation.

As a candidate Galactic PeVatron, SS~433 is expected to produce a high-energy neutrino flux comparable with the gamma-ray one.
In particular, in the multi-pion regime one expects the production rates of neutrinos and gamma rays to scale as $2 \, E_{\nu}^2 Q_{\nu}(E_{\nu}) \approx 3 \, E_{\gamma}^2 Q_{\gamma}(E_{\gamma})$, where $E_{\gamma} \approx 2 E_{\nu}$.
Consequently, in the TeV range one can expect the neutrino flux from SS~433 to be on the order of $E_{\nu}^2\Phi_{\nu}^{SS433}(E_{\nu}) \approx 10^{-12}-10^{-13} \, \rm TeV \, cm^{-2} \, s^{-1}$.
It is interesting to explore how upcoming neutrino observatories located in strategic positions to observe Galactic objects, such as KM3NeT~\citep{KM3_NeT_ARCA}, will be able to constrain the possible hadronic emission from the wind bubble inflated by SS~433.
Following \cite{KM3_NeT_ARCA}, we considered the 12-year discovery potential of KM3NeT-ARCA for a point-like source with $E^{-2}$ spectrum: $\Phi_{\rm DP}^{PL} \approx 5 \cdot 10^{-10} \, \rm GeV^{-1} \, cm^{-2} \, s^{-1}$. Moreover,  we assumed $\sigma_{\rm PSF} = 0.2\deg$ as a representative uncertainty for track-like events~\citep{Ambrosone_2024}. This allowed us to estimate the discovery potential of the extended SS~433 wind bubble as
\begin{equation}
    E_{\nu}^2 \Phi_{\rm DP}^{\rm SS~433}(E_{\nu}) \approx E_{\nu}^2 \Phi_{\rm DP}^{PL} (E_{\nu}) \sqrt{\frac{\sigma_{\rm PSF}^2 + \sigma_{\rm src}^2}{\sigma_{\rm PSF}^2}} \approx 10^{-12} \, \rm TeV \, cm^{-2} \, s^{-1},
\end{equation}
where we assumed $\sigma_{\rm src} = r_{39} \approx 0.32\deg$.
This suggests that during its first decade KM3NeT might already be able to detect the bubble of SS~433 if the actual flux from the source is at the level of the most optimistic realizations shown in Fig.~\ref{Fig: Modelling SS433}.

The radio domain can be an interesting alternative window for testing particle acceleration taking place at the TS of ULX bubbles. 
In the following, we present an analytic argument providing an order-of-magnitude estimate of the low-energy leptonic emission from primary electrons accelerated in these systems. 

We began our estimation assuming a primary electron population at the shock 
$dn_{e}/d\gamma_e \sim \gamma_e^{2-s} {\rm exp}[-\gamma_em_ec^2/E_{\rm max, e}]$.
We assumed $s=4$ as prescribed by test-particle DSA, and we estimated the maximum energy by comparing the acceleration with the radiative-loss timescale of electrons. 
The latter is governed by synchrotron losses, which dominate over the inverse Compton cooling on the CMB, since $U_B \simeq 4 \cdot 10^{-12}\, B_1^2 \, {\rm erg \, cm}^{-3}  \simeq 10\, U_{\rm CMB}$, where $B_1$ is the magnetic field expressed in units of $10 \, \mu \rm G$ (typical value for the shocked wind bubble; see Eq.~\ref{eq:B}). The synchrotron-loss timescale then reads 
\begin{equation}
\tau_{\rm syn, e} \approx 0.2~{\rm Gyr} \, B_{1}^{-2} \gamma_{e,3}^{-1}, 
\label{eq:tau_syn}
\end{equation}
and the maximum electron energy achieved is
\begin{equation}
    E_{\rm max,e} \approx 0.5 \, V_{w,9}^{2/(3-\delta)} \, l_{c,-1}^{(1-\delta)/(3-\delta)} B_1^{-\delta/(3-\delta)} \, \rm PeV.
\end{equation}
Unless different background photon fields or stronger magnetic fields are present, ULXs can also be electron PeVatrons. Therefore, for our qualitative argument we assume $E_{\rm max,e} = 1 \, \rm PeV$.

As we are interested in estimating the flux at radio wavelengths, we focused on relatively low electron energies. 
Using $\nu^* =1~{\rm GHz} \, \nu^*_{9}$ as the observing frequency, we estimated the typical Lorentz factor of radiating electrons. Following \cite{R&L1986}, we obtain $\gamma_{e}^* \approx 6 \cdot 10^3 \, \nu^{* 1/2}_{9}  B^{-1/2}_1$ (or $E_{e}^* \approx 3 \,  \nu^{* 1/2}_{9} B^{-1/2}_1 \,  \rm GeV$). Since the cooling timescale of these electrons (see Eq.~\ref{eq:tau_syn}) is much longer than their advection timescale,  $f_{e,\rm ts}$ is, with good approximation, spatially constant throughout the whole wind bubble. This also suggests a radial morphology of the radio emission.

Using a delta-function approximation for the single-particle synchrotron emissivity~\citep{Ghisellini2013}, $j_S(\nu) = j_0 \delta(\nu-\nu_S)$, where $\nu_S =  e B \gamma_e^2/(2 \pi m_e c)$ and $j_0 = (4/3) \sigma_T c U_B \gamma_e^2$, we may write the synchrotron flux at $\nu$ as 
\begin{equation}
    F_S(\nu) \approx   j_S(\nu)  \frac{{\rm d}n_e}{{\rm d}\gamma_e} \frac{{\rm d}\gamma_e}{{\rm d}\nu} \frac{\frac{4}{3} \pi R_{\rm fs}^3}{4 \pi D_L^2},
\end{equation}
where the emitting volume is taken to be equal to the volume of the shocked wind with $R^3_{\rm fs}\gg R^3_{\rm ts}$, and $D_L$ is the luminosity distance of the source. 
Assuming, as in the case of protons, that non-thermal electrons take a fraction, $\xi_e \approx 10^{-2} \xi_{\rm CR,}$ of the ram pressure at the shock, the synchrotron flux can be rewritten as 
\begin{align}
    \nu^* F_S(\nu^*) & \approx \frac{\sigma_T}{6 \pi m_e^2 c^3} \frac{E^*_e U_B}{D_L^2} \frac{\xi_e \Dot{M} V_{w} }{\Lambda_e} \frac{R_{\rm fs}^3}{R_{\rm ts}^2} \nonumber \\
    & \approx 10^{-14} \, \frac{E^*_e B^2_1}{D_{L,3.74}} \frac{\xi_{e,-3} \Dot{M}_{-6} V_{w,9} }{\Lambda_{e,1.32}} \frac{R_{\rm fs,2}^3}{R_{\rm ts,1}^2} \, \rm erg \, \rm cm^{-2} \, s^{-1},
\end{align}
where $\Lambda_e = {\rm ln}[E_{\rm max, e}/(m_ec^2)] \approx 21$ and $D_L$ is normalized to $10^{3.74} {\rm pc} \approx 5.5 \, \rm kpc$, the distance of SS~433. 
Interestingly, our order-of-magnitude prediction is in good agreement with the NVSS observation of the bubble at around 1.4 GHz~\citep{Condon_NVSS}. 
This could be a hint for {a substantial contribution of} the wind bubble {to the spherical component of the} nebula W50 surrounding SS~433~\citetext{see \citealp{Churazov} for a detailed presentation of a model including both an isotropic and collimated wind}.

{Although the association of the spherical part of W50 and at least part of its radio emission with a supernova remnant cannot be discarded,}
the qualitative argument presented in this subsection {is a plausible alternative and} illustrates that a leptonic population in the wind bubble can already be probed through radio observations. A more detailed comparison {with} existing radio data would require the full solution of Eq.~\eqref{Eq: transport} and account {for} the secondary electron population injected through the decay of charged pions resulting from $pp$ interactions. We leave this investigation for a follow-up study.

\section{ULXs as Galactic PeVatrons}\label{sec:pevatrons}

The maximum energies achievable at the TS of ULX bubbles makes them extremely interesting PeV accelerators. For this reason, we also explore the role of ULXs as sources of Galactic CRs in the knee region.

The possible contribution of ULXs to the CR flux at Earth could be relevant if at least one of the following two scenarios is true: 
{(1) there are enough ULXs in our Galaxy to allow their escaping CR flux to diffuse and fill the whole Galactic volume; (2) the Earth is close enough to a ULX to allow CRs to reach us without escaping the Galaxy.}

We begin by considering the first scenario. According to \cite{Kovlakas_ULX_census}, the average number of {observable} ULXs in a galaxy can be inferred from astronomical observations as follows:
\begin{equation}
    \mathcal{N}_{\rm ULX}^{\rm {(obs)}} = 0.45^{+0.06}_{-0.09} \times \frac{\rm SFR}{\rm M_{\odot} \, yr^{-1}} + 3.3^{+3.8}_{-3.2} \times \frac{M_{\star}}{10^{12} \rm M_{\odot}},
\end{equation}
where SFR is the star-formation rate and $M_{\star}$ is the stellar mass of the Galaxy. Using the typical SFR and stellar mass of the Milky Way \citep{Galactic_SFR}, the expected number of {observable} ULXs in our Galaxy is on the order of unity.
{We highlight that the total number of ULXs can be larger than the observed one as these sources are geometrically beamed. In particular, assuming a standard opening angle of $\theta_{f} = 20\deg$ for the funnel, the fraction of observable ULXs will be rescaled with the factor $\mathcal{F}= (1- {\rm cos} \, \theta_{f}) \approx 0.06$
\citetext{see also \citealp{Lasota23} and \citealp{Kayanikhoo25} for comparable estimates}.}

Bearing in mind that the expected number of {observable} ULXs in the Galaxy should be extremely limited, {while the actual number could be larger,} we evaluate the total number of ULXs necessary to contribute to the CR flux at PeV energies (i.e., at the knee of the CR spectrum) in the following.
In order to do so, we assumed {isotropic} diffusive transport in the Galaxy with the Galactic diffusion coefficient $D_{\rm Gal}(E) = D_0 E_{\rm GeV}^{{{\delta_G}}} = 3 \cdot 10^{28} \, E_{\rm GeV}^{{{\delta_G}}} \, \rm cm^2 \, s^{-1} $, {where ${{\delta_G}}$ is the energy dependence of the Galactic diffusion coefficient} \citetext{see, e.g., \citealp{AMS2016} and \citealp{Genolini2017}}.
The energy flux of CRs in stationary conditions in the Galaxy can be written as
\begin{equation}
\label{Eq: flux_earth}
    E^2 \Phi_{\oplus}(E)=\frac{H_{\rm Gal}^2}{D_{\rm Gal}(E)} \frac{c}{4 \pi} \mathcal{N}_{\rm ULX} \frac{E^2 J_{\rm esc}(E)}{V_{\rm Gal}}
,\end{equation}
where $\mathcal{N}_{\rm ULX}$ is the {total} number of ULXs in the Galaxy, $V_{\rm Gal}=2 \pi H_{\rm Gal} R_{\rm Gal}^2$ is the Galactic volume, $R_{\rm Gal}$ the Galactic radius, {and $H_{\rm Gal} \approx 4 \, \rm kpc$ is the Galactic halo size \citep[see, e.g.,][]{Evoli2018}}.
Under the assumption of $s=4$ and neglecting the specific shape of the exponential high-energy cut-off, the escaping CR power from a ULX, $E^2 J_{\rm esc}(E)$, can be written as
\begin{equation}
\label{Eq: escaping}
    E^2 J_{\rm esc}(E) \approx \frac{3 \xi_{\rm CR}}{2 \Lambda} \Dot{E}_{\rm kin} 
,\end{equation}
where $\xi_{\rm CR}$ is the ratio between the cosmic ray pressure and ram pressure at the wind shock, for which we expect something on the order of $1-10 \%$, and $\Lambda \approx {\rm ln}(E_{\rm max}/E_{\rm inj}) \approx 14$. One can see that Eq.~\eqref{Eq: escaping} was obtained assuming an $E^{-2}$ spectral shape extending up to the maximum energy and neglecting the specific form of the high-energy cut-off.

Knowing that $E^2 \Phi_{\oplus}(1 {\rm PeV}) \approx 10^{-4} \, \rm GeV \, cm^{-2} \, s^{-1} \, sr^{-1}$~\citetext{see, e.g., \citealp{Recchia_2024}, and references therein}, it is possible to invert Eq.~\eqref{Eq: flux_earth} and write the total number of ULXs as
\begin{equation}
    \mathcal{N}_{\rm ULX} \approx {20}  \cdot \, [E^2 \Phi_{\oplus}]_{-4} D_{28.5} H_{\rm Gal,0.7}^{-1} R_{\rm Gal,1.2}^2 \xi^{-1}_{\rm CR, -1} \Lambda_{1.2} {\Delta_G}\Dot{E}_{\rm kin,{39}}^{-1},
\end{equation}
where the kinetic power is computed assuming the benchmark values of $\Dot{M}$ and $V_w$ (see Table~\ref{Table: SS433}){, and ${\Delta_G ={10^{6 \delta_G -2}}}$}.
This suggests that, while remaining consistent with the order-of-magnitude prediction of their {observable number} in the Galaxy~\citetext{(see also \citealp{Mapelli2010} and \citealp{Prestwich2013})}, Galactic ULXs might {meet the energy requirements to} substantially contribute to the observed CR flux at the knee. 
Their relative reduced number {would indeed be} consistent with their non detection, ${\mathcal{N}_{ULX}^{\rm (obs)} \approx \mathcal{F} \mathcal{N}_{ULX} }$, in that all sources might simply be misaligned with favorable lines of sight or highly obscured. 
Assuming that ULXs are uniformly distributed in the circular Galactic disk, the average distance between them would be $d_{\rm ULX} \approx 128\,  R_{\rm Gal}/ [45 \pi \sqrt{\mathcal{N}_{\rm ULX}}] \approx 3 R_{\rm Gal,1.2} \mathcal{N}_{\rm ULX,1.3}^{-1/2} {\rm kpc,}$ making our estimate plausible but somehow close to a critical threshold where its applicability may become questionable as $d_{\rm ULX} \lesssim H_{\rm Gal}$. 
In fact, as the average distance between ULXs is expected to be on the order of a couple of kiloparsecs, it is reasonable to expect that the flux at Earth might be dominated by one, or at most two, nearby sources. This leads us to consider the second scenario.

Even though there are no ULXs detected in the Milky Way according to the observational definition of $L_X \gtrsim 10^{39} \rm erg \, s^{-1}$, there may be super-Eddington accreting sources with their funnels pointing away from our line of of sight, making them hidden in X-rays. 
In fact, according to their current state of accretion, SS~433 \citep{King2023} and Cygnus-X3 \citep{Valedina_Cygnus_X3} are considered (hidden) Galactic ULXs, {and other objects, such as 4FGL J1405.1-6119 \citep{Saavedra2023}, might soon be added to the list}.
As currently there are no other closer ULX candidates in our Galaxy, it is reasonable to assume that SS~433 is the closest one. 
However, its distance ($\sim 5.5$~kpc), which is larger than the inferred height of the Galactic magnetized halo, implies that for isotropic diffusion, most of its CR flux would diffuse away from the Galaxy before reaching us.    
{Therefore, unless peculiar transport conditions enhance the escaping flux of CRs from SS~433, this source seems unlikely to dominate the flux at the knee. 
On the other hand, if a highly obscured and undetected ULX were relatively close to Earth, its contribution to the CR spectrum at the knee could be substantial.}

\section{Conclusions}\label{sec:conclusions}

In this work, we explored the power of diffusive shock acceleration in the wind bubbles excavated by ULXs. 
We did so by solving the space and energy-dependent cosmic-ray transport equation, and particularly by focusing on the particle acceleration at the wind termination shocks of these bubbles. 
We studied the maximum energy of these accelerators and discussed the associated multi-messenger radiation in terms of gamma rays, high-energy neutrinos, and radio.
We finally explored the possible role of ULXs as Galactic cosmic-ray sources. 

We found that for a wide range of typical parameters, ULXs can accelerate protons up to several peta-electronvolts, while the most powerful sources could energize up to tens of peta-electronvolts. 
This suggests that ULX wind bubbles can be considered as a new class of PeVatrons. 

We applied our model to the super-Eddington accreting Galactic source SS~433 where we found that the maximum energy can be as high as 3-8 PeV and the associated wind bubble could be already detectable in gamma rays as a diffuse GeV component. 
The bubble emission might also explain the flux of gamma-rays observed by LHAASO at energies $>100$~TeV as well as the change in the gamma-ray source morphology taking place at around that energy. In particular, we argued that below 100 TeV the collimated jets dominate the emission, whereas at higher energies, where the inverse Compton is expected to fade down, the hadronic bubble emission can be observed.

We noticed that in \cite{LHAASO_BH_2024}, the hadronic interpretation is supported by a spatial coincidence of the emission (>100 TeV) with a giant molecular cloud. As the real position of such a cloud with respect to the geometrical boundaries of the bubble is uncertain, we leave a detailed study of the bubble-cloud modeling to a follow-up investigation.
However, we point out that if the cloud was embedded inside the forward shock radius, the system might already be qualitatively described by our scenario S2. On the other hand, if the cloud was situated at larger distances than the forward shock, a proper diffusion model would be applied to the escaping particles to infer the actual cloud luminosity.

We estimated the high-energy neutrino flux from the wind bubble of SS~433, and we concluded that a neutrino observatory such as KM3NeT could detect a flux or constrain some realizations of our model with about ten years of observations. Moreover, our order-of-magnitude estimate of the radio flux at 1~GHz (due to synchrotron radiation of accelerated electrons in the bubble)  is comparable with the total radio flux in the same energy band as measured from the nebula W50 that surrounds SS~433. 
This result led us to speculate on the possible ULX-bubble nature of {the spherical part of W50,} instead of interpreting it as a supernova remnant. 
{In particular, the presence of a mildly relativistic disk wind \citep{2021MNRAS.506.1045M} might substantially increase former estimates \citep{Chi-SNR-ss433} of its impact on the overall structure of the nebula.}

We finally considered the possible role of ULXs as Galactic PeVatrons in contributing to the cosmic-ray spectrum at the knee. 
According to an observational scaling based on the star-formation rate, the number of {observable} ULXs expected in our Galaxy should be on the order of unity, {while,} according to our estimates, about {20} of these objects would be required {to saturate the CR flux at the knee}. 
{Since ULXs are geometrically beamed, the two numbers are consistent with each other as all ULXs might be not favorably aligned with our line of sight. However, such a small number of ULXs in the Galaxy would result in one, or at most two, nearby objects dominating the observed flux.}
With our current knowledge, given the distance of the nearest Galactic ULX candidate, SS~433, the conclusion shall be that ULXs are unlikely to dominate the CR flux at the knee. 
{However}, we argue that {in addition to} Cygnus X-3 {and} SS~433,{ in agreement with our estimate, there could be other obscured ULXs in the Galaxy}. 
We conclude that ULXs cannot be discarded as dominant cosmic-ray sources at the knee, as some other undetected, or unidentified, ULXs could be hosted in our Galaxy {within a few kiloparsecs of the Earth}. 

Ultra-luminous
X-ray sources are powerful sources powered by super-Eddington accretion onto stellar-mass compact objects. Our investigation highlighted their multi-messenger aspects and showed that they can be sources of PeV cosmic rays, high-energy gamma rays, radio, and neutrinos in the Galaxy.

\begin{acknowledgements}
      E.P. is grateful to R. Amato for insightful discussions and to G. Israel for useful comments on the manuscript. 
      E.P. is also grateful to V. Bosh-Ramon, S. Recchia, C. Pinto and L. Crosato Menegazzi for valuable comments that allowed improving the first version of the manuscript, and E. Amato, G. Morlino, N. Bucciantini, D. Caprioli, C. Evoli, A. Condorelli and A.Ambrosone for insightful discussions and comments. 
      M.P. and G.V. would like to thank Université Paris Cité, where this project was conceived, for their hospitality.  E.P. and S.G. were supported by Agence Nationale de la Recherche (grant ANR-21-CE31-0028).
      M.P. acknowledges support from the Hellenic Foundation for Research and Innovation (H.F.R.I.) under the ``2nd call for H.F.R.I. Research Projects to support Faculty members and Researchers'' through the project UNTRAPHOB (Project ID 3013). G.V. also acknowledges support by the Hellenic Foundation for Research and Innovation (H.F.R.I.) under the ``3rd Call for H.F.R.I. Research Projects to support Postdoctoral Researchers'' through the project ASTRAPE (Project ID 7802).  
\end{acknowledgements}

\bibliographystyle{aa}
\bibliography{paper_v1} 

\appendix

\section{Gamma rays from $pp$ interactions}
\label{Appendix}
We compute the production spectra of high-energy gamma rays ($E_\gamma \ge 100$~GeV)
following the approach described in \cite{Kelner_pp}:
\begin{align}
    \frac{dN_{\gamma}(E_{\gamma})}{dE_{\gamma}} & = c \int_0^{R_{\rm fs}} dr \, 4 \pi r^2 n(r) \int_{p_{\rm min}}^{\infty} \frac{dp}{E_p(p)} \, 4 \pi p^2 \, \times \nonumber \\ 
    & f(r,p) \, \sigma_{\rm inel}(p) \, F_{\gamma}\left(\frac{E_{\gamma}}{E_p(p)},E_p(p)\right),
    \label{Eq: Gamma HE}
\end{align}
where $p_{\rm min}= \sqrt{E_{\gamma}^2/c^2 - m_p^2c^2}$, while $\sigma_{\rm inel}$ and $F_{\gamma}$ are given in Eqs. (79) and (58) of \cite{Kelner_pp}. 
Low energy ($E_{\gamma}<100 \, \rm GeV$) gamma rays are computed following the delta-function approximation described in \cite{Kelner_pp}:
\begin{equation}
    \frac{dN_{\gamma}(E_{\gamma})}{dE_{\gamma}} = 2  \int_0^{R_{\rm fs}} dr \, 4 \pi r^2  \int_{E_{\rm min}}^{\infty} \frac{q_{\pi}(r,E_{\pi})}{\sqrt{E_{\pi}^2-m_{\pi}^2c^4}} dE_{\pi},
\end{equation}
where $E_{\rm min}= E_{\gamma} + m_{\pi}^2 c^4/4 E_{\gamma}$, $m_{\pi}$ is the pion mass and the production rate of $\pi^0$ mesons reads
\begin{align}
    q_{\pi}(E_{\pi}) = & \frac{\Tilde{n}c}{K_{\pi}} n(r) \, \sigma_{\rm inel}\left[\left(m_p c+\frac{E_{\pi}}{c K_{\pi}}\right)^2 - m_p^2c^2\right] \times \nonumber \\
    & 4 \pi p^2 \, \frac{dp}{dE_{p}} \, f\left(r,\left[\left(m_p c+\frac{E_{\pi}}{c K_{\pi}}\right)^2 - m_p^2c^2\right]\right)
\end{align}
where $\Tilde{n}$ and $K_{\pi}$ are free parameters guaranteeing continuity with the high-energy branch (Eq.~\eqref{Eq: Gamma HE}).

\end{document}